\documentclass{elsart}

\usepackage{graphicx}% Include figure files
\usepackage{dcolumn}% Align table columns on decimal point
\usepackage{amsmath,amsfonts,amssymb}% popular packages from the American Mathematical Society
\usepackage{bm}% Bold Math package
\newcommand{\intinfty}[0]{\int \limits_{-\infty}^{\infty}}

\begin{document}

%\preprint{}
\begin{frontmatter}
\title{An Algorithm for the Continuous Morlet Wavelet Transform}

\author{Richard B\"ussow}
 \address{Institute of Fluid Mechanics and Engineering Acoustics, Berlin University of Technology, Einsteinufer 25, 10587 Berlin}
\ead{richard.buessow@tu-berlin.de}
             %  but any date may be explicitly specified

\begin{abstract}
This article consists of a brief discussion of the energy density over time or 
frequency that is obtained with the wavelet transform. Also an efficient
algorithm is suggested to calculate the continuous transform with the Morlet wavelet. \\
The energy values of the Wavelet transform are compared with the power spectrum of the Fourier transform. Useful definitions for power spectra are given. \\
The focus of the work is on simple measures to evaluate the transform with the Morlet wavelet in an efficient way. The use of the transform and the defined values is shown in some examples. 
\end{abstract}
\end{frontmatter}

\section{Introduction}
 The wavelet transform is a method for time-frequency analysis.  The
 application for  acoustic signals can be found in several publications. These 
 publications deal for example with the analysis of dispersive waves
 \cite{onsay,kishimoto},  source or damage localization
 \cite{gaulident,junsheng},  investigation of system parameters
 \cite{hayashi,ta} or active control \cite{berry98wavelet}.  A comparison of 
 the short time Fourier transform and the wavelet transform is done by Kim
 et.al.  \cite{kimkim}.  It is found that the continuous wavelet transform CWT
 of  acoustic signals is a promising method to obtain the time - frequency
 energy  distribution of a signal.\\
 Another sort are wavelets for spatial transforms, that are an effective tool for
 damage localisation\cite{rucka,spatialwav}. \\
The aim of this publication is to present an algorithm for the continuous
wavelet  transform CWT with the Morlet wavelet. Caprioli et al. state in  
\cite{caprioli} that "The high computational complexity (of the CWT) is  not a
serious worry for today's computer power; still it might be an obstacle for  an
"on line" application." Alternatives  to the CWT are characterised by low 
computational complexity, but the best results are obtained by the CWT. The 
presented algorithm implements simple measures to reduce the computational 
complexity of the CWT. An implementation in Java is written and is open  source
and freely available online\footnote{\tt http://www.tu-berlin.de/fb6/ita}. \\ 
Nevertheless it should be mentioned that the Wigner approximation is also very useful in the given context. A recent publication is for example \cite{cohen}. Other methods, that are adapted to the analysis of dispersive waves are discussed in \cite{yoonkim,fulop}. \\
This publication starts with a discussion of the energy values that are obtained by the wavelet transform. It follows a description of an efficient algorithm to evaluate the transform.  

\section{Morlet Wavelet Transform}
The theoretical background of the wavelet transform can be found in textbooks 
\cite{Mallat98wavelet,wavelets}. Only the used definitions are stated. The
wavelet  transform with the wavelet $\psi$ of a signal $y(t)$ is defined as
\begin{equation}
W_\psi^y (a,b) = \frac{1}{\sqrt{c_\psi |a|}} \intinfty y(t) \psi\left(\frac{t - b}{a} \right) \: dt,
\label{cwtdef}
\end{equation}
where  $a$ is called dilatation and $b$ translation parameter. The Morlet 
wavelet $\psi$ which is sometimes called Gabor wavelet is given by
\begin{equation}
\psi(t) =  e^{-\beta \frac{t^2}{2}}e^{j \omega_0 t}
\label{psimorlet}
\end{equation}
and $c_\psi = \sqrt{\pi/ \beta}$. The values $\beta=\omega_0^2$ and $\omega_0$
are  defined in a particular application so that the admissibility condition is 
valid \cite{wavelets}. The function (\ref{psimorlet}) is called mother wavelet. \\ 
The Morlet wavelet is common for the time frequency  analysis of acoustic
signals. For a comparison of different so called mother wavelets see Schukin et.
 al. \cite{schukin}.
\subsection{Comparing Fourier and wavelet transform}
\label{secenpow}
When using the Fourier transform one usually develops the spectrogram. The wavelet transform has scaling factors. The analog to the spectrogram is the scalogram defined as
\begin{equation}
 |W_\psi^y (a,b)|^2.   
\end{equation}
The scalogram is a measure of the energy distribution over time shift $b$ and scaling factor $a$ of the signal. It holds that the energy $E_y$ of a signal $y$ is
\begin{equation}
E_y = \intinfty |y(t)|^2 \: dt = \intinfty \intinfty |W_\psi^y (a,b)|^2 \frac{da \: db}{a^2}.
\label{enerwav}
\end{equation}
If instead of the scaling factor $a$ the frequency value $f=1/a$ is used, the
value $f$ is only the {\it real} frequency if $\omega_0 = 2 \pi$. \\  
It follows with $ da = \frac{da}{df} \: df = - \frac{1}{f^2} \: df $ that
\begin{equation}
E_y = \intinfty \intinfty |W_\psi^y (f,b)|^2 \:df \:db.
\label{energydoubleint}   
\end{equation}
It is possible to divide this total energy into an energy density over time and over frequency. This is achieved by one integration over frequency or time. 
The energy density over time is defined by
\begin{equation}
E_t(b) = \intinfty |W_\psi^y (f,b)|^2 \:df.
\end{equation}
The energy density over frequency, or the energy density spectrum is defined by
\begin{equation}
E_f(f) = \intinfty |W_\psi^y (f,b)|^2 \:db.
\end{equation}

\subsubsection{Power signals}
The value $|W_\psi^y (f,b)|^2$ is sometimes called Morlet power spectrum, but here the term energy density is used. A Comparision of the values obtained with the Fourier transform and the wavelet transform is, for example, given in \cite{perrier95wavelet}. A power signal is characterized by
\begin{equation}
P = \frac{1}{T} \intinfty |x(t)|^2 \: dt < \infty.
\end{equation}
 If the wavelet transform is applied to a power signal one recognises that for
 higher  frequencies the energy density is lower, as shown in the examples in
 section  \ref{examples}. To achieve a better match, Shyu \cite{shyu} proposes a
  modified equal amplitude wavelet power spectrum that is given by
\begin{equation}
\mbox{MPS} = \frac{f}{c_\psi} \: |W_\psi^y (f,b)|^2. 
\label{mps}
\end{equation}
The above definition corresponds more closely to power spectra obtained
with the discrete Fourier transform. This can be explained since $P= \frac{1}{T} E$
and $c_\psi a$ is the effective length of the wavelet so $T_{eff} = c_\psi/f$ is
the value to scale the energy density. Since one is usually familiar with power
spectra, the above definition (\ref{mps}) is useful but a problem is that
\begin{equation}
P \ne \intinfty \intinfty \mbox{MPS} \:df \:db. 
\end{equation}
This fact can be compared with the windowed Fourier transformation. If the
window like  for example, the flat top is defined so that the peak value is
constant\footnote{Which is strictly true only if $  f_n = T/(2 n \pi)$, but the
window  should minimize the deviation.}, it follows that the effective
noise band  width is altered by the windowing function. The Morlet wavelet
transform can be interpreted as a windowed Fourier transformation with a
frequency dependent  window size. 
\paragraph{Alternative definition of the power spectrum} In the following a  new
definition of the power spectrum is given. It is defined analogous to the  
discrete Fourier transform, so that the power can be calculated by summation of
the power values. The discrete Fourier
transform is a filter with the bandwidth $\Delta f$, so each
value of the power spectrum is the integral over $\Delta f$.  The power spectrum
 that is defined with the energy values of the discrete continuous Morlet
 wavelet transform is
\begin{equation}
P_i = \frac{\Delta f_i}{T}  E_f(f_i) . 
\label{pimor}
\end{equation}
Given the shape of the Morlet wavelet, the only sensible scaling of the
frequency values is logarithmic. With this prerequisite the values $P_i$ follow the same
tendency as equation (\ref{mps}). Using equation (\ref{pimor}) the total power
is given by $P = \sum_i P_i$. \\
From equation (\ref{pimor}) and $\Delta t = T/N_t$, where $N_t$ is the number of
time values, it follows that
\begin{equation} 
P_d (i,j) = \frac{1}{T} \, \Delta f_i \, \Delta t \, |W_\psi^y (f_i,b_j)|^2 = \frac{\Delta f_i}{N_t} |W_\psi^y (f_i,b_j)|^2 
\label{pddef}
\end{equation}
is a reasonable definition for the power spectrum over time. An example for a power signal is given in section \ref{powersec}.

\subsubsection{Energy signals}
However, the discrete Fourier Transform is usually the best method to work with
power signals. The wavelet transform in contrast is suited
to work with energy signals, for which 
\begin{equation}
E = \intinfty |x(t)|^2 \: dt < \infty
\end{equation}
is true. It is useful to use the value $E_f$ for comparing Fourier and
Wavelet spectra in this case. This is done by calculating 
\begin{equation}
E_f(i) = P_i \frac{T}{ \Delta f_i}
\label{effourier} 
\end{equation}
from the power spectrum obtained with the Fourier transform. An example of an
energy  signal is given in \ref{energysec}.

\section{CWT Algorithm}
\label{cwtalgo}
%The section begins with a brief recapitulation of algorithms that are in use and can be found in the literature. It follows a description of features that are implemented in a new algorithm. 
%A decision against a commercial implementation was taken, since they are normally not open source and at least the version used in MathWorks Wavelet Toolbox does not have the desired features that will be discussed in the following sections. 
Usually the continuous wavelet transform is implemented in a direct algorithm.
The starting  point is a program written by Hayashi \cite{suzuki}, where the
integral  in equation (\ref{cwtdef}) is implemented by a simple summation over 
the elements. Typical values for the scaling parameter $a$ are 20 to 40, where 
$a_n= \alpha^n$, with $\alpha=\sqrt[4]{2}$. The time shift parameter is $b_m =
m \Delta t$,  where $0 \le m \le N$. \\
%Please that $\omega_0$ is not defined as above, but this only adds a factor. \\
The algorithm to calculate the wavelet transform of a discrete signal $y_i$ of 
the length $N$ is for each $a_n$:
\begin{enumerate}
\item Calculate $\psi_i$, of the length $2 N$, $b_m = m \Delta t$ with 
$\omega_0=\sqrt{2} \pi /\Delta t $.
\item Shift $\psi$ about $m$ : $ \sum_i^N y_i \cdot \psi_{N-m+i}$
\end{enumerate}
The frequency is given by $f=\frac{\omega_0 }{2 \pi a_n}$.  
% which is 
%\begin{equation}
%f=2^{-\frac{n+2}{4}} f_s.
%\label{freqagu}
%\end{equation}

\subsection{Improved algorithm} 
The continuous wavelet transform is very inefficient compared with the discrete 
wavelet transform. A brief description of a new algorithm follows which
introduces  ideas that are crucial for the high efficiency of the discrete
wavelet  transform DWT. Besides the following improvements the program
calculates  the corresponding frequency values and energy or power distributions
 that are defined in section \ref{secenpow}.
\subsubsection{Effective Compact support} \label{effcompactsupp}
The outer parts of the wavelet $|t| \gg 0$ have no significant contribution  to
the result because of the exponentially decaying envelope term $e^{-\beta
t^2/2}$  which acts like a window, known as the windowed Fourier transform.  Due
to the limited floating point precision of a computer, the outer
parts will not contribute to the integral in equation (\ref{cwtdef}). \\
 An a-priori estimation of the effective wavelet length  is made here. For $t=0$
 the envelope has its maximum which is unity. The length is chosen so,  that for
 all $|t| > |t_{max}| $ the envelope is smaller than the precision. For a  64
 bit floating point number this means that
\begin{eqnarray}
e^{-\beta t_{max}^2/2} &=& 2.22 \cdot 10^{-16} \nonumber \\
|t_{max}|  &=& 8.49 / \sqrt{\beta}. 
\label{tmax}
\end{eqnarray}
Strictly the outer parts can still contribute to the integral (\ref{cwtdef})
especially if $y$  is a transient function, but in practise even a weaker
definition  does not show a significant effect.\\
The concept of neglecting the outer parts is called effective compact support. 
Note that to develop the transform the wavelet is scaled with $t/a$. A general 
straight forward implementation is not possible, due to the discrete time
values. 
This is a reason why the discrete wavelet transform uses scaling 
values that are defined by $a_n=2^n$, where $n$ is an integer value. \\
The  actual implementation works in the following way. The length of the wavelet
that represents the lowest frequency $f_{min}$ is chosen so that $T/ \Delta t = 2^N$.  
The mixed approach is that the wavelet length is halved for each doubling of 
the frequency, so for each value $f_n=2^n f_{min}$, and left otherwise unchanged. For 
all frequency values $f_i$ for which $f_n=2^n f_{min} < f_i<2^{n+1}
f_{min}=f_{n+1}$
holds, it follows that the support of the wavelet is even longer than
$t_{max}$. The reason for this effect is that also $t_{max}$ is scaled and so
for a constant length and a higher frequency the support of the mother wavelet
is longer.

\subsubsection{Frequency dependent time shift} \label{freqshift}
Due to the argument $\frac{t - b}{a}$ of the wavelet $\psi$ in (\ref{cwtdef})
the transform can be interpreted as a convolution of a signal $y$ and a wavelet $\psi$
\begin{equation}
y \ast \psi = \intinfty y(t) \psi(t - \tau) \: dt.   
\end{equation}
With the scaled argument $t/a$ and $\tau = b/a$  equation (\ref{cwtdef})
follows. The algorithm  that is outlined in section \ref{cwtalgo} does actually 
not set the translation parameter $b$, but $b/a$. It is the highest precision  
of the continuous translation parameter $b$ with discrete time values. But it
is inefficient, since the whole wavelet and its time resolution are scaled with
$a$. The resolution depends on the wavelength. If one increases the time shift
for  lower frequencies this does not result in a less accurate resolution. Like 
for the effective compact support the translation parameter is adjusted for each
 doubling of the frequency.\\
The implementation uses the frequency dependent length of the a period
$T_\lambda = 1/f$.  The time shift parameter $b$ is defined relative to the
period length $T_\lambda$ and an arbitrary dyadic factor $2^{n_b}$ in a way that
$T_\lambda/2^{n_b} \ge b > T_\lambda/2^{n_b+1}$ is valid. The details can be
found in the Java-code. \\ 
In some cases the values given above are not fully applicable. For example for  
frequencies $f > f_s/2^{n_b}$, with the sampling frequency $f_s$, the smallest possible
shift is $\Delta t$ which can be bigger than $T_\lambda/2^{n_b}$, but it is the
highest possible precision. The  
default value is $2^{n_b}=16$, which results in a high resolution where a difference
 to a higher resolution is not recognizable. With this measure a reduced
 computational  effort is achieved. \\
 Together with the compact support it is
 possible  to use an amount of 200 scaling factors instead of 20 and wait a few 
 seconds for the transform to be calculated. 
 
\subsubsection{Frequency dependent truncated boundaries}
One may interpret $y(t)$ as an infinite signal of which a part of the length  
$T$ is known. The convolution of the wavelet over the signal results in a  
considerable violation of the admissibility condition since all the parts  of
the wavelet that lie outside the known signal must vanish. In practise this 
results in invalid data at the boundaries. When working with the compact  
support this is avoided if the algorithm guaranties that the whole wavelets  
always fits in the signal. Since the compact support of the wavelet will be 
shorter for high frequencies this results in a frequency dependent truncation. 
This measure is also implemented in \cite{jordan}.

\subsection{Fast Morlet wavelet transform}
A fast implementation is described in \cite{jordan} and often called fast Morlet
 wavelet transform. The convolution in equation (\ref{cwtdef}) is implemented by
  building the DFT of the wavelet and the analysed function, multiplying them
  and  building the inverse Fourier transform. Since the DFT is calculated with
  the  fast Fourier transform (FFT) algorithm, the computational effort is
  greatly  reduced compared with \cite{suzuki}.

\subsection{Computational effort} \label{compeff}
This section will compare the order of the amount of operations that are needed
for each implementation. Since the amount of operations depends on the
actual implementation such an investigation is beyond the scope of this investigation. 
\begin{description} 
\item[AGU-implementation] For the implementation \cite{suzuki} the signal of
length $N$  has to be
multiplied  $N$ times with the wavelet. For $M$ scaling factors the order of the
amount of operations is $O(M \times N^2)$.
\item[Fast Morlet Wavelet transform] The FFT has a computational effort of
$O(log(N) N)$, if $N$ is a power of two. It follows that the order of operation 
for the fast Morlet wavelet transform is $O(M \times log(N) N)$.
\item[New algorithm] According to section \ref{effcompactsupp} the number of
points of the wavelet $N_w$ is independent of the signal length $N$. For a  
given frequency value $f_i$ the number of points is inversely dependent on
the frequency $N_w \sim 1/f_i$. \\ According to section
\ref{freqshift} the translations parameter is also inversely
dependent on the frequency $b_i \sim 1/f_i$. The wavelet
has to be multiplied $N/b_i$ times with the signal. Since $N_w<N$ the amount of
multiplications is $N_w N/b_i$. It follows that the order of the amount of 
multiplications for a given frequency value is $O(N)$. The resulting overall 
effort is $O(M \times N)$.
\end{description}
\subsubsection{Benchmark}
The consumed time for the transform depends on many other factors. It is only
an  indication for the amount of operation that actually have to be performed. 
Important factors that effect the computation time is the systems memory and the
implementation itself. \\ 
The benchmark compares the standard Morlet wavelet transform (st), the fast  
Morlet wavelet transform (fast) and the described new Algorithm (new) in a typical application. \\
The standard Morlet wavelet transform is implemented with Matlab's {\tt conv()} 
function, that implements the convolution with a Matlab's {\tt filter()}
function.  Matlab's profiler shows that $99.3\%$ of the computational time is 
used by the function {\tt conv()} which is optimized. \\
The fast Morlet wavelet transform is implemented by a self-written convolution 
function that uses Matlabs highly optimized {\tt fft()} function. The {\tt
fft()}-function itself chooses the presumable best algorithm for the
application.  The FFT algorithm needs a number of points that are an integer  
power of two $2^n$. This can be problematic if one wants to adjust the  length
of a window so that it fits the discrete best case frequencies $f_n = T/(2 n
\pi)$.  This defiancy does not apply to the continuous wavelet transform. \\
The described algorithm is implemented with Java and called from Matlab. It is 
not optimised and it is known that Java programs of this sort usually suffer  a
50-fold performance degradation \cite{javaslow} compared to native Fortran  or
similar applications. The intention of the implementation is to follow the
guidelines for software engineering: Build the functionality first and do the  
optimization afterwards. The advantages of Java are that it is platform
independent,  has an easy to use interface to Matlab and no special or
commercial  libraries are needed. 
\begin{table}
\label{bench}
\begin{center}
\begin{tabular}{|l|*{10}{|c}|}
\hline
 N & $2^8$ & $2^9$ & $2^{10}$ & $2^{11}$ & $2^{12}$ & $2^{13}$ & $2^{14}$ & $2^{15}$ & $2^{16}$ & $2^{17}$ \\
 \hline
st in s&     
    0.51 &
    1.90 &
    7.85 &
    27.3 &
   106 &
   417 &
  1602 & 
  6445 & &  \\
$ \frac{T \times 10^{-6}}{N^2} $ &
    0.79 &
    0.72 &
    0.75 &
    0.65 &
    0.63 &
    0.63 &
    0.60 &
    0.60 & &  \\
fast in s&
    0.16 &
    0.27 &
    0.49 &
    0.80 &
    1.38 &
    3.15 &
    6.76 &
   16.48 & &  \\

$ \frac{T \times 10^{-3}}{N \log(N)} $ &
    0.39 &
    0.29 &
    0.23 &
    0.17 &
    0.16 &
    0.15 &
    0.14 &
    0.14 & &   \\

\hline
new in s& 
    0.07 &
    0.23 &
    0.34 &
    0.45 &
    0.83 &
    1.37 &
    2.62 &
    4.83 &
    9.12 &
   18.16  \\
$ \frac{T \times 10^{-3}}{N} $& 
    0.27 &
    0.45 &
    0.33 &
    0.22 &
    0.20 &
    0.17 &
    0.16 &
    0.15 &
    0.14 &
    0.14  \\
\hline
\end{tabular}
\caption{Computation time for three different algorithms: standard (st), fast and new Morlet wavelet transform and for different sizes.}
\end{center}
\end{table}
Table \ref{bench} shows the computational time for the same $M=200$ frequency  
values and a maximum frequency $f_{max}=f_s/8$. It is assumed that measurements 
are often taken with a portable computer, so a Samsung X20 notebook is chosen, 
equipped with Pentium M CPU at 1.73 GHz and 1GB main memory. The parameters for 
the new transform were chosen such that the results are identical, a time 
shift parameter of $T_\lambda/4$ and $t_{max}=5.6s$. The results show a good
repeatability and the same tendency on other computers. \\
The order of operations that is derived in section \ref{compeff} is validated with the
normalised computational time $T/N^2$ for the standard, $T/(N \log(N))$ for the fast
and
$T/N$ for the new algorithm. If the computational time would solely depend on
the amount of operations the normalised computational time should be constant. \\ 
 When using the standard and fast algorithms the size of the memory limited  the
test to a maximum number of $2^{15}$ points. With the new algorithm it was
possible to calculate the transform also for $2^{18}$ and $2^{19}$ in a time of 
 $37.25s$ and $69.16s$. These values also indicate that the estimation of a  
 linear increasing number of operations is valid, since $ \frac{T \times
 10^{-3}}{N} $  is again $0.14$. For a typical sampling frequency of $f_s =
 48kHz$  this means that it is possible to analyse around $10s$.
\section{Examples}
\label{examples}
The capabilities of the wavelet transform and the described algorithm are  
demonstrated in the examples below. 
\subsection{Power signal}
\label{powersec}
\begin{figure}
\includegraphics[width=0.9 \textwidth]{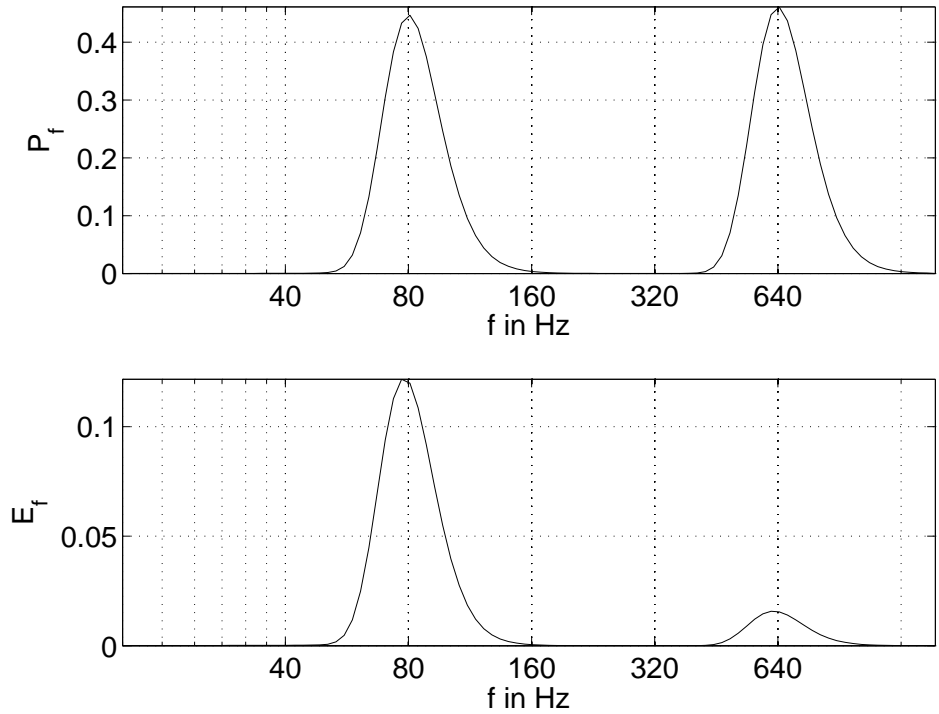}
\caption{Power density spectrum equation (\ref{pimor}) and energy density
spectrum of  equation (\ref{exzweisin}) built with the Morlet wavelet}
\label{zweisinEfP}
\end{figure}

\begin{figure}
\includegraphics[width=0.9 \textwidth]{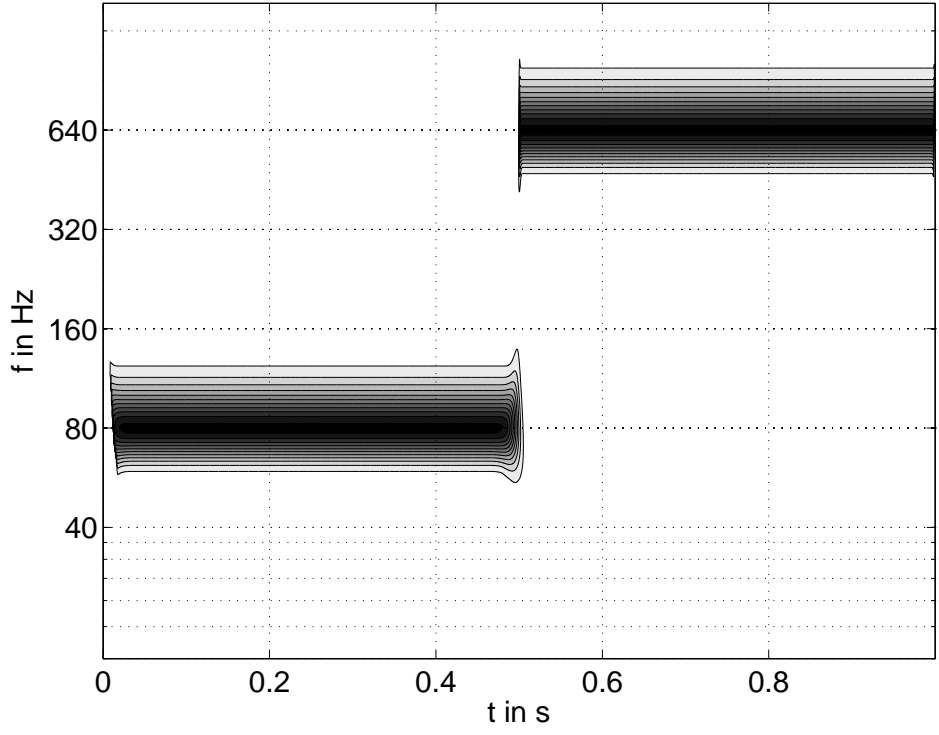}
\caption{Contour plot of the power density spectrum  over time, equation
(\ref{pddef}),  of the example  equation (\ref{exzweisin}) built with the Morlet wavelet}
\label{zweisinPd}
\end{figure}

A typical power signal is the sine function. Consider the  example with a changing frequency  
\begin{equation}
y(t) = \left\{ \begin{array}{l}
A \sin(2 \pi f_1 t) \mbox{ for } t < 1/2 \\
A \sin(2 \pi f_2 t) \mbox{ for } t > 1/2,
\end{array} \right.
\label{exzweisin}
\end{equation}
with $\Delta t=2^{-13}$, $f_1 = 80 Hz$, $f_2 = 640 Hz$ and $A=4$. The power
density spectrum equation (\ref{pimor}) is obtained with the fast Morlet
wavelet  transform and the presented algorithm for 100 frequency values.  The
results are identical to those of the other algorithms and plotted in figure  
\ref{zweisinEfP}. \\
The peaks are very broad which is expected and due to the $e^{-\beta/2 t^2}$   
windowing term. The peak at $f_1=80 Hz$ is lower than that at $640Hz$, which  is
due to the truncation described above and
that is plotted in figure \ref{zweisinPd}. The peak value does not match the
amplitude. This is due to the fact that the total power is approximately
correct. Since the peak is very broad only the peak value or the total can power
can be correct. The total power is $P=\sum_i P_i = 8.1$ which corresponds to the
correct value of $P=1/2 A^2 = 8$. Using the equal amplitude wavelet power
spectrum that is given in equation
(\ref{mps}) and performing an integration over time results in a good approximation
of the peak value. \\
The energy density spectrum is plotted in the bottom part of figure \ref{zweisinEfP}, showing an
unusual representation of the example equation (\ref{exzweisin}). \\
A contour plot of the power density spectrum over time is given in figure \ref{zweisinPd}. 
Note that the high frequency component $f_2$ is more accurately localized in
the time domain.\\

\subsection{Energy signal}
\label{energysec}
\begin{figure}
\includegraphics[width=0.9 \textwidth]{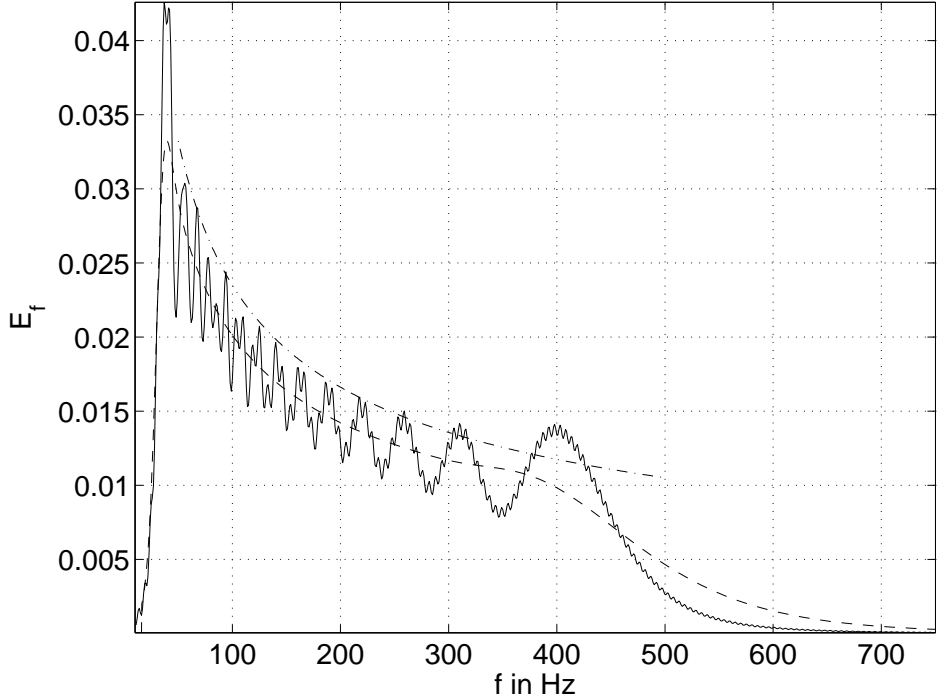}
\caption{Energy density spectrum of equation (\ref{exchirp}) built with the
Morlet wavelet (solid), the Fourier transform (dashed) and $1/\sqrt{\omega}$ (dash-dotted)}
\label{chirpef}
\end{figure}

\begin{figure}
\includegraphics[width=0.9 \textwidth]{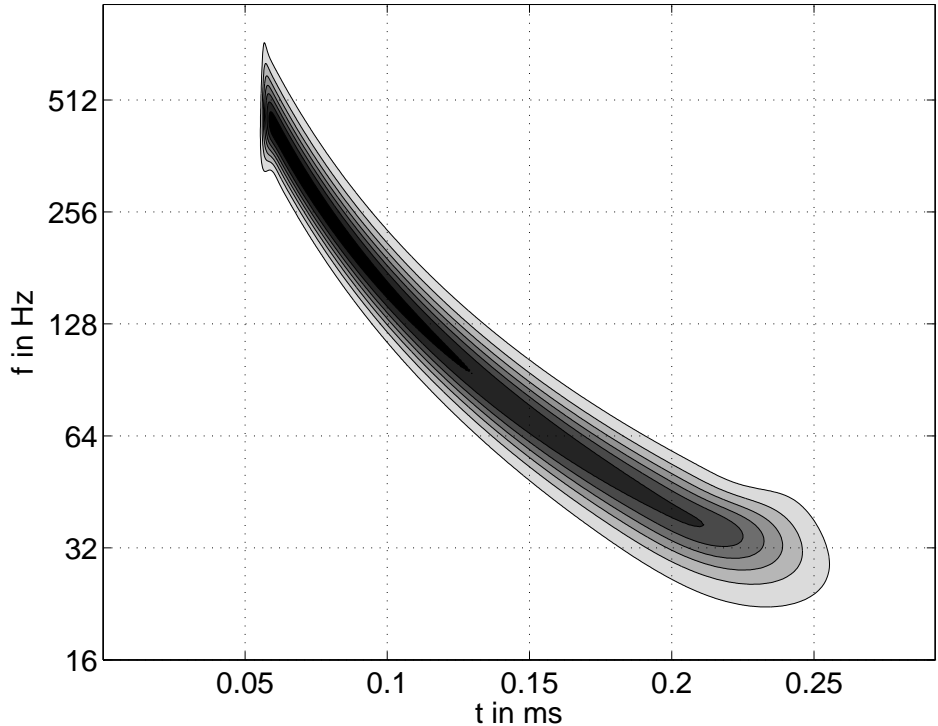}
\caption{Contour plot of the energy density spectrum over time of equation (\ref{exchirp})}
\label{chirpscalo}
\end{figure}

The tested signal is 
\begin{equation}
y(t) = \frac{\sin(a/t)}{t} \mbox{ for } t_{min}< t < t_{max},
\label{exchirp}
\end{equation}
thereby, the parameters are set to $a=10$, $1/\Delta t=f_s=2^{12}$, $t_{min} =
\sqrt{a/(2 \pi f_{max})}$, $f_{max}=f_s/8$  and $t_{max}=.25s$. \\
The energy is plotted in figure \ref{chirpef}. The solid curve is calculated 
with the discrete Fourier transform (DFT) and equation (\ref{effourier}). The 
Fourier transform of function (\ref{exchirp}) can be obtained by the
convolution  of the Bessel-function $J_0$ which is the Fourier transform of   $
\frac{ \sin (a/t)}{t}$ and the sinc-function which is the Fourier transform of
the rectangular window of the size $T=t_{max}-t_{min}$. 
\begin{equation}
F\{y(t)\} = \frac{\pi}{2} J_0(2 \sqrt{a \omega}) \ast \frac{\sin(T \omega)}{\omega}. 
\end{equation}
The small fluctuations in figure \ref{chirpef} correspond to the sinc-function
and  the high and long to the Bessel-function.\\
The dashed curve is calculated with the Morlet wavelet transform. This curve 
follows more clearly the theoretical $1/\sqrt{\omega}$ dependence that is  
plotted as the dash dotted curve next to the others. \\ 
A contour plot of the energy density spectrum over time is given in figure 
\ref{chirpscalo}. It shows good agreement with the actual frequency of this 
function which is given by $\omega(t) = a/t^2$. 
\section{Concluding remarks}
The study shows that it is possible to obtain an efficient transform with an
algorithm that is restricted to the values that are numerically significant. This
efficiency results in less
computational effort and less memory consumption. Nevertheless,  for
applications that are time critical it will be worth to work on a optimized  
version in a native language and to apply optimisation. Since the programme  is
tested and open source, it will facilitate the programming of such an optimised version. \\
The examples and definitions of energy and power values should help with the 
interpretation of the values obtained with the wavelet transform.   
\bibliographystyle{elsart-num}
\bibliography{lite}
\end{document}